\documentclass[12pt,a4paper]{article}
\usepackage{amsmath}
\usepackage{latexsym}
\usepackage[margin=2cm,noheadfoot]{geometry}

\allowdisplaybreaks[1]
\input{tcilatex}
\begin{document}

\title{$CP$--Violation in $B_{q}$ Decays and Final State Strong Phases}
\author{Fayyazuddin\thanks{%
fayyazuddins@gmail.com} \\
National Centre for Physics \&\\
Department of Physics, Quaid-i-Azam University, Islamabad.}
\date{{\footnotesize PACS: 12.15.Ji, 13.25.Hw, 14.40.Nd}}
\maketitle

\begin{abstract}
Using the unitarity, $SU(2)$ and $C$-invariance of hadronic interactions,
the bounds on final state phases are derived. It is shown that values
obtained for the final state phases relevant for the direct $CP$-asymmetries 
$A_{CP}(B^{0}\rightarrow K^{+}\pi ^{-},K^{0}\pi ^{0})$ are compatiable with
experimental values for these asymmetries. For the decays $B^{0}\rightarrow
D^{(\ast )-}\pi ^{+}$ $(D^{(\ast )+}\pi ^{-})$ described by two independent
single amplitudes $A_{f}$ and $A_{\bar{f}}^{\prime }$ with differnt weak
phases ($0$ and $\gamma $) it is argued that the $C$-invariance of hadronic
interactions implies the equality of the final state phase $\delta _{f}$ and 
$\delta _{\bar{f}}^{\prime }$. This in turn implies, the $CP$-asymmetry $%
\frac{S_{+}+S_{-}}{2}$ is determined by weak phase ($2\beta +\gamma )$ only
whereas $\frac{S_{+}-S_{-}}{2}=0.$ Assuming factorization for tree graphs,
it is shown that the $B\rightarrow D^{(\ast )}$ form factors are in
excellent agreement with heavy quark effective theory. From the experimental
value for $\left( \frac{S_{+}+S_{-}}{2}\right) _{D^{\ast }\pi },$ the bound $%
\sin (2\beta +\gamma )\geq 0.69$ is obtained and $\left( \frac{S_{+}+S_{-}}{2%
}\right) _{D_{S}^{\ast -}K^{+}}\approx -(0.41\pm 0.08)\sin \gamma $ is
predicted. For the decays described by the amplitudes $A_{f}\neq A_{\bar{f}}$
such as $B^{0}\longrightarrow \rho ^{+}\pi ^{-}:$ $A_{\bar{f}}$ and $%
B^{0}\longrightarrow \rho ^{-}\pi ^{+}:A_{f}$ where these amplitudes are
given by tree and penguin diagrams with differnt weak phases, it is shown
that in the limit $\delta _{f,\bar{f}}^{T}\rightarrow 0,r_{f,\bar{f}}\cos
\delta _{f,\bar{f}}=\cos \alpha $ and $\frac{S_{\bar{f}}}{S_{f}}=\frac{%
S+\Delta S}{S-\Delta S}=-\frac{\sqrt{1-C_{\bar{f}}^{2}}}{\sqrt{1-C_{f}^{2}}}%
. $
\end{abstract}

\section{ Introduction}

The CP asymmetries in the hadronic decays of B and K mesons involve strong
final state phases. Thus strong interactions in these decays play a crucial
role. The short distance strong interactions effects at quark level are
taken care of by perturbative QCD in terms of Wilson coefficients. The CKM
matrix, which connects the weak eigenstates with mass eigenstates, is
another aspect of strong interactions at quark level. In the case of semi
leptonic decays, the long distance strong interaction effects manifest
themselves in the form factors of final states after hadronization. Likewise
the strong interaction final state phases are long distance effects. These
phase shifts essentially arise in terms of S-matrix which changes an 'in'
state into an 'out' state viz. 
\begin{equation}
|f\rangle _{in}=S|f\rangle _{out}=e^{2i\delta _{f}}|f\rangle _{out}
\label{01}
\end{equation}

In fact, the CPT invariance of weak interaction Lagrangian gives for the
weak decay $B(\bar{B})\rightarrow f(\bar{f})$ 
\begin{equation}
\bar{A}_{\bar{f}}\equiv _{out}\langle \bar{f}| \mathcal{L}_{w}|\bar{B}%
\rangle=\eta_{f}e^{2i\delta_{f}}A_{f}{\ast}  \label{02}
\end{equation}

Taking out the weak phase $\phi$, the amplitude $A_{f}$ can be written as 
\begin{equation}
A_{f}=e^{i\phi}F_{f}=e^{i\phi}e^{i\delta_{f}}|F_{f}|  \label{02a}
\end{equation}

Then Eq. $\eqref{02}$ implies 
\begin{equation*}
\bar{A}_{\bar{f}} = e^{-i\phi}e^{2i\delta_{f}}F_{f}^{\ast} = e^{-i\phi}F_{f}
\end{equation*}

It is difficult to reliably estimate the final state strong phase shifts. It
involves the hadronic dynamics. However, using isospin, C-invariance of
S-matrix and unitarity, we can relate these phases. In this regard,
following cases are of interest:

Case (i): The decays $B^{0}\rightarrow f, \bar{f}$ described by two
independent single amplitudes $A_{f}$ and $A_{\bar{f}}^{\prime}$ with
different weak phases: 
\begin{align*}
A_{f} &=\langle f\left| \mathcal{L_{W}}\right| B^{0}\rangle = e^{i\phi}F_{f}
= e^{i\phi}e^{i\delta_{f}} \bigl|F_{f}\bigr| \\
A_{\bar{f}}^{\prime} &=\langle \bar{f}\left| \mathcal{L_{W}^{\prime}}\right|
B^{0}\rangle =e^{i\phi^{\prime }} F_{\bar{f}}^{\prime} = e^{i\phi^{\prime
}}e^{i\delta^{\prime}_{\bar{f}}} \bigl|F_{\bar{f}}^{\prime}\bigr|
\end{align*}

where the states $|\bar{f}\rangle $ and $|f\rangle $ are C conjugate of each
other such as states $D^{(*)-}\pi ^{+}(D^{(*)+}\pi ^{-})$, $%
D_{s}^{(*)-}K^{+}(D_{s}^{(*)+}K^{-})$, $D^{-}\rho ^{+}(D^{+}\rho ^{-})$.

For case (i), there is an added advantage that the decay amplitudes $A_{f}$
and $A_{\bar{f}}$ are given by tree graphs. Assuming factorization for tree
amplitudes, it is shown that the form factors $f_{0}^{B-D} (m_{\pi}^{2})$, $%
A_{0}^{B-D^{\ast}} (m_{\pi}^{2})$, $f_{+}^{B-D} (m_{\rho}^{2})$ obtained
from the experimental branching ratios are in excellent agreement with Heavy
Quark Effective Theory (HQET). Hence factorization assumption is
experimentally on sound footing for these decays.

Case (ii): The weak amplitudes $A_{f} \ne A_{\bar{f}}$, 
\begin{align*}
A_{f} &=\langle f\left| \mathcal{L_{W}}\right| B^{0}\rangle = \left[%
e^{i\phi_{1}}F_{1f} + e^{i\phi_{2}}F_{2f} \right] \\
A_{\bar{f}} &=\langle \bar{f}\left| \mathcal{L_{W}}\right| B^{0}\rangle = %
\left[e^{i\phi_{1}}F_{1\bar{f}} + e^{i\phi_{2}}F_{2\bar{f}} \right]
\end{align*}

as is the case for the following decays, 
\begin{align*}
B^{0} \rightarrow \rho^{-}\pi^{+}(f): A_{f}&, \quad B^{0} \rightarrow
\rho^{+}\pi^{-}(\bar{f}): A_{\bar{f}} \\
B_{s}^{0} \rightarrow K^{\ast -}K^{+}&, \quad B_{s}^{0} \rightarrow K^{\ast
+} K^{-} \\
B^{0} \rightarrow D^{\ast -}D^{+}&, \quad B^{0} \rightarrow D^{\ast +}D^{-}
\\
B_{s}^{0} \rightarrow D_{s}^{\ast -}D_{s}^{+}&, \quad B_{s}^{0} \rightarrow
D_{s}^{\ast +}D_{s}^{-}
\end{align*}

The $C-$ invariance of S-matrix gives $S_{\bar{f}}=S_{f}$ which implies 
\begin{equation*}
\delta _{f}=\delta _{\bar{f}}^{\prime },\qquad \delta _{1f}=\delta _{1\bar{f}%
},\qquad \delta _{2f}=\delta _{2\bar{f}}
\end{equation*}

\section{Unitarity and Final State Strong Phases}

The time reversal invariance gives 
\begin{equation}
F_{f} = _{out}\langle f|\mathcal{L}_{W}|B\rangle = _{in}\langle f|\mathcal{L}%
_{W}|B\rangle^{\ast}  \label{03}
\end{equation}

where $\mathcal{L}_{W}$ is the weak interaction Lagrangian without the CKM
factor such as $V_{ud}^{\ast}V_{ub}$. From Eq. $\eqref{03}$, we have 
\begin{align}
F_{f}^{\ast} =& _{out}\langle f|S^{\dagger}\mathcal{L}_{W}|B\rangle  \notag
\\
=& \sum_{n} S_{nf}^{\ast}F_{n}  \label{04}
\end{align}

It is understood that the unitarity equation which follows from time
reversal invariance holds for each amplitude with the same weak phase. Above
equation can be written in two equivalent forms:

\begin{enumerate}
\item Exclusive version of Unitarity \cite{1,2}\newline
Writing 
\begin{equation}
S_{nf}=\delta _{nf}+iM_{nf}  \label{05}
\end{equation}%
we get from Eq (\ref{04}) , 
\begin{equation}
\text{Im}F_{f}=\frac{1}{2}\sum_{n}M_{nf}^{\ast }F_{n}  \label{06}
\end{equation}

where $M_{nf}$ is the scattering amplitude for $f\rightarrow n$ and $F_{n}$
is the decay amplitude for $B\rightarrow n$. In this version, the sum is
over all allowed exclusive channels. This version is more suitable in a
situation where a single exclusive channel is dominant one. To get the final
result, one uses the dispersion relation. In dispersion relation two
particle unitarity gives dominant contribution. From Eq.(\ref{06}), using
two particle unitarity, we get \cite{1}, 
\begin{equation}
Disc\text{ }F(B\rightarrow f^{\prime })\approx \frac{1}{16\pi s}%
\int_{-\infty }^{0}M_{f^{\prime }f}^{\ast }F(B\rightarrow f)dt  \label{6a}
\end{equation}%
where $t=-2\vec{p}^{2}(1-\cos \theta )$, $\left| \vec{p}\right| \approx 
\frac{1}{2}\sqrt{s}.$ Eq.$\left( \ref{6a}\right) $ is especially suitable to
calculate rescattering corrections to color suppressed $T$-amplitude in
terms of color favored $T$-amplitude as for example rescattering correction
to color suppressed decay $B^{0}\rightarrow \pi ^{0}\bar{D}^{0}(f)$ in terms
of dominant decay mode $B^{0}\rightarrow \pi ^{+}D^{-}(f)$. Before using two
particle unitarity in this form, it is essential to consider two particle
scattering processes.

$SU(3)$ or $SU(2)$ and $C$-invariance of $S$-matrix can be used to express
scattering amplitudes in terms of two amplitudes $M^{+}$ and $M^{-}$ which
in terms of Regge trajectories are given by \cite{3,4,5}%
\begin{eqnarray}
M^{(+)} &=&P+f+A_{2}=-C_{P}\frac{e^{-i\pi \alpha _{p}(t)/2}}{\sin \pi \alpha
_{p}(t)/2}\left( s/s_{0}\right) ^{\alpha (t)}  \notag \\
&&-2C_{\rho }\frac{1+e^{-i\pi \alpha (t)}}{\sin \pi \alpha (t)}\left(
s/s_{0}\right) ^{\alpha (t)}  \label{6b} \\
M^{(-)} &=&\rho +\omega =2C_{\rho }\frac{1-e^{-i\pi \alpha (t)}}{\sin \pi
\alpha (t)}\left( s/s_{0}\right) ^{\alpha (t)}  \label{6c}
\end{eqnarray}%
For linear Regge trajectories, using exchange degeneracy, we have \ 
\begin{eqnarray}
\alpha _{\rho }(t) &=&\alpha _{A_{2}}(t)=\alpha _{\omega }(t)=\alpha
_{f}(t)=\alpha \left( 0\right) +\alpha ^{\prime }t,  \notag \\
\alpha _{p}(t) &=&\alpha _{p}(0)+\alpha _{p}^{\prime }(t),  \notag \\
C_{f} &=&C_{\omega };\text{ }C_{A_{2}}=C_{\rho };\text{ }C_{\omega }=C_{\rho
}  \label{6d}
\end{eqnarray}%
We take $\alpha _{0}\approx 1/2$, $\alpha ^{\prime }\approx 1$GeV$^{-2},$ $%
\alpha _{p}(0)\approx 1,\alpha _{p}^{\prime }\approx 0.25$GeV$^{-2}$. Using $%
SU(3)$ and taking $\gamma _{\rho D^{+}D^{-}}=\gamma _{\rho K^{+}K^{-}},$ we
get $C_{\rho }=\gamma _{\rho \pi ^{+}\pi ^{-}}\gamma _{\rho
K^{+}K^{-}}=\gamma _{\rho \pi ^{+}\pi ^{-}}\gamma _{\rho D^{+}D^{-}}=\frac{1%
}{2}\gamma _{0}^{2},$ $\gamma _{0}=\gamma _{\rho \pi ^{+}\pi ^{-}};\gamma
_{0}^{2}\approx 72\cite{3}.$ Hence for $\pi ^{+}D^{-}$ or $\pi ^{-}K^{+}$
scattering we get 
\begin{eqnarray}
M &=&M^{(+)}+M^{(-)}=iC_{P}e^{bt}(s/s_{0})  \notag \\
&&+2\gamma _{0}^{2}ie^{\alpha ^{\prime }(\ln (s/s_{0})-i\pi
)t}(s/s_{0})^{1/2}  \label{6e}
\end{eqnarray}%
where $b=\alpha _{P}^{\prime }\ln (s/s_{0})$

For $\pi ^{0}\bar{D}^{0}\rightarrow \pi ^{+}D^{-}$, $\pi
^{0}K^{0}\rightarrow \pi ^{-}K^{+}$%
\begin{equation}
M=\pm \sqrt{2}M^{(-)}=\pm i2\sqrt{2}C_{\rho }\frac{e^{-i\pi \alpha (t)/2}}{%
\cos \alpha (t)/2}(s/s_{0})^{\alpha (t)}  \label{6f}
\end{equation}%
From Eq.(\ref{6a}) and (\ref{6f}) with the use of dispersion relation, we
obtain 
\begin{eqnarray}
A(B^{0} &\rightarrow &\pi ^{0}\bar{D}^{0})_{FSI}=\frac{\sqrt{2}\gamma
_{0}^{2}(1-i)}{16\pi }\frac{A(B^{0}\rightarrow \pi ^{+}D^{-})}{\ln \left( 
\frac{m_{B}^{2}}{s_{0}}\right) +i\pi /2}\frac{1}{\pi }%
\int_{(m_{B}+m_{D})^{2}}^{\infty }\frac{ds}{s-m_{B}^{2}}(s/s_{0})^{\alpha
(t)}  \notag \\
&=&-\sqrt{2}\epsilon A(B^{0}\rightarrow \pi ^{+}D^{-})e^{i\theta }
\label{6g}
\end{eqnarray}%
We get $\epsilon \approx 0.06,\theta \approx 33^{\circ }$ by putting $%
s\approx m_{B}^{2}$ in $\ln (s/s_{0})$. Now $A(B^{0}\rightarrow \pi
^{+}D^{-})=T.$ Hence with rescattering correction \cite{6} 
\begin{eqnarray}
A(B^{0} &\rightarrow &\pi ^{0}\bar{D}^{0})=-\frac{1}{\sqrt{2}}C-\sqrt{2}%
\epsilon Te^{i\theta }  \notag \\
&=&-\frac{C}{\sqrt{2}}\left[ 1+\frac{\epsilon }{b}e^{i\theta }\right]
\label{6h}
\end{eqnarray}%
where $2b=C/T.$ Hence the final state phase shift $\delta _{C}$ for the
color suppressed amplitude induced by the final state interaction is given
by 
\begin{equation}
\tan \delta _{C}=\frac{\epsilon /b\sin \theta }{1+\epsilon /b\cos \theta }%
\rightarrow \delta _{C}\approx 8^{\circ }  \label{6i}
\end{equation}%
with $b\approx 0.174,$ which we get from 
\begin{equation}
\frac{\Gamma (B^{0}\rightarrow \pi ^{+}D^{-})}{\Gamma (B^{+}\rightarrow \pi
^{+}\bar{D}^{0})}=\frac{1}{(1+2b)^{2}}\approx 0.55\pm 0.03  \label{6j}
\end{equation}%
For $B^{0}\rightarrow \pi ^{0}K^{0},$ the color suppressed $T$-amplitude
with rescattering correction is given by 
\begin{equation}
-\frac{1}{\sqrt{2}}C+\sqrt{2}\epsilon Te^{i\theta }=-\frac{1}{\sqrt{2}}C%
\left[ 1-\frac{\epsilon }{b}e^{i\theta }\right]  \label{6k}
\end{equation}%
where $2b=C/T\approx 0.37$ \cite{7}. Hence $\delta _{C}$ generated by the
final state interaction is given by 
\begin{equation}
\tan \delta _{C}=\frac{-\epsilon /b\sin \theta }{1-\epsilon /b\cos \theta }%
\rightarrow \delta _{C}\approx -8^{\circ }  \label{6l}
\end{equation}

To conclude: The scattering amplitude $M\left( s,t\right) $ for the two
particle final state obtained in eq.$\left( 13\right) $ is used in the
unitarity equation to generate the final state strong phase by rescattering
for the color suppressed tree amplitude.

\item Inclusive version of Unitarity [2]\newline
This version is more suitable for our analysis. For this case, we write Eq. %
\eqref{04} in the form%
\begin{equation}
F_{f}^{\ast }-S_{ff}^{\ast }F_{f}=\sum_{n\neq f}S_{nf}^{\ast }F_{n}
\label{07a}
\end{equation}
\end{enumerate}

Parametrizing S-matrix as \ $S_{ff}\equiv S=\eta e^{2i\Delta }$\cite{5}, $%
0\leq \eta \leq 1,$ we get after taking the absolute square of both sides of
Eq.(\ref{07a})%
\begin{equation}
\left\vert F\right\vert ^{2}\left[ (1+\eta ^{2})-2\eta \cos 2(\delta
_{f}-\Delta )\right] =\sum_{n,n^{\prime }}F_{n}S_{nf}^{\ast }F_{n^{\prime
}}^{\ast }S_{n^{\prime }f}  \label{08}
\end{equation}

The above equation is an exact equation. In the random phase approximation
[2], we can put 
\begin{align}
\sum_{n^{\prime },n\neq f}F_{n}S_{nf}^{\ast }F_{n^{\prime }}S_{n^{\prime
}f}=& \sum_{n\neq f}|F_{n}|^{2}|S_{nf}|^{2}  \notag \\
=& \bar{|F_{n}|^{2}}(1-\eta ^{2})  \label{09}
\end{align}

We note that in a single channel description \cite{5,8}: 
\begin{equation*}
(Flux)_{in}-(Flux)_{out}=1-|\eta e^{2i\Delta }|^{2}=1-\eta ^{2}=\text{%
Absorption}
\end{equation*}

The absorption takes care of all the inelastic channels. \newline
Similarly for the amplitude $F_{\bar{f}}$, we have 
\begin{equation}
F_{\bar{f}}^{\ast} - S^{\ast}_{\bar{f}\bar{f}}F_{\bar{f}} = \sum_{\bar{n}
\neq \bar{f}} S^{\ast}_{\bar{n}\bar{f}}F_{\bar{n}}  \label{10}
\end{equation}

The C-invariance of S-matrix gives: 
\begin{align}
S_{fn} =& \langle f|S|n\rangle = \langle f|C^{-1}CSC^{-1}C|n\rangle  \notag
\\
=& \langle \bar{f}|S|\bar{n}\rangle = S_{\bar{f}\bar{n}}  \label{11}
\end{align}

Thus in particular C-invariance of S-matrix gives \ \ 
\begin{equation}
S_{\bar{f}\bar{f}}=S_{ff}=\eta e^{2i\Delta }  \label{12}
\end{equation}

Hence from Eq. $\left( \ref{08}\right) $, using Eqs. $(\ref{09}-\ref{12})$,
we get 
\begin{equation}
\frac{1}{1-\eta ^{2}}[(1+\eta ^{2})-2\eta \cos 2(\delta _{f,\bar{f}}-\Delta
)]=\rho ^{2},\bar{\rho}^{2}  \label{13}
\end{equation}

where 
\begin{equation}
\rho ^{2}=\frac{\overline{\bigl|F_{n}\bigr|}^{2}}{\bigl|F_{f}\bigr|^{2}}%
,\qquad \bar{\rho}^{2}=\frac{\overline{\bigl|F_{\bar{n}}\bigr|}^{2}}{\bigl|%
F_{\bar{f}}\bigr|^{2}}  \label{14}
\end{equation}%
From Eq.(\ref{13}), we get 
\begin{equation}
\sin (\delta _{f,\bar{f}}-\Delta )=\pm \sqrt{\frac{1-\eta ^{2}}{4\eta }}%
\left[ \rho ^{2},\bar{\rho}-\frac{1-\eta }{1+\eta }\right] ^{1/2}  \label{15}
\end{equation}%
The maximum value for $\rho ^{2},\bar{\rho}^{2}$ is 1 and the minimum value
for them is $\frac{1-\eta }{1+\eta }.$ Hence we get the following bounds:%
\begin{eqnarray}
\frac{1-\eta }{1+\eta } &\leq &\rho ^{2},\bar{\rho}^{2}\leq 1  \notag \\
0 &\leq &\delta _{f,\bar{f}}-\Delta \leq \theta  \notag \\
-\theta &\leq &\delta _{f}-\Delta \leq 0  \label{16a} \\
\theta &=&\sin ^{-1}\sqrt{\frac{1-\eta }{2}}  \label{16b}
\end{eqnarray}%
From now on, we will confine our self to positve square root in Eq,(\ref{15}%
).

The strong interaction parameter $\Delta $ and $\eta $ in the above bounds
can be obtained from the scattering amplitude $M(s,t)$ given in Eq.(12)
obtain from Regge pole analysis. The $s-$wave scattering amplitude $f$ is
given by \ 
\begin{equation}
f\approx \frac{1}{16\pi s}\int_{-s}^{0}M(s,t)  \label{30}
\end{equation}%
For the scattering amplitude $M=M^{+}+M^{-}$ relevant for $\pi ^{+}D^{-},\pi
^{-}K^{+}$ and $\pi ^{+}\pi ^{-}$, we obtain from Eq.(\ref{30}) using
Eq.(12) 
\begin{eqnarray}
f &=&f_{P}+f_{\rho }=\frac{1}{16\pi s}\frac{iC_{P}}{b}\left( \frac{s}{s_{0}}%
\right) +2\frac{\gamma _{0}^{2}}{16\pi }\frac{1}{\ln (s/s_{0})-i\pi }%
(s/s_{0})^{-1/2}  \label{31a} \\
&=&\left[ 
\begin{array}{c}
\text{0.12}i\text{ + (-0.08+0.08}i\text{)} \\ 
\text{0.17}i\text{ +(-0.08+0.08}i\text{)} \\ 
\text{0.16}i\text{+(-0.16}\pm \text{0.16}i\text{)}%
\end{array}%
\right]  \label{31b}
\end{eqnarray}%
where we have used $s\approx m_{B}^{2}\approx (5.27)^{2}$ GeV$^{2}$. For $%
C_{P}$ we have used the values of reference [2] whereas for $C_{\rho
}=\gamma _{\rho \pi ^{+}\pi ^{-}}\gamma _{\rho K^{+}K^{-}}=\gamma _{\rho \pi
^{+}\pi ^{-}}\gamma _{\rho D^{+}D^{-}}=\frac{1}{2}\gamma _{0}^{2}$ and $%
C_{\rho }=\gamma _{\rho \pi ^{+}\pi ^{-}}\gamma _{\rho \pi ^{+}\pi
^{-}}=\gamma _{0}^{2}\approx 72$ \ for $\pi D,\pi K$ and $\pi \pi $
respectively.

\ \ \ \ \ Using the relation $S=\eta e^{2i\Delta }=1+2if,$ where $f$ is
given by $Eq.(33)$, the phase shift $\Delta ,$ the parameter $\eta $ and the
phase angle $\theta $ can be determined. One gets 
\begin{equation*}
\pi ^{+}D^{-}(\pi ^{-}D^{+}):\Delta \approx -7^{\circ },\eta \approx
0.62,\theta \approx 26^{\circ }
\end{equation*}%
\begin{eqnarray}
\pi ^{-}K^{+}\text{ or }\pi ^{0}K^{0} &:&\Delta \approx -9^{\circ },\eta
\approx 0.52,\theta \approx 29^{\circ }  \notag \\
\pi ^{+}\pi ^{-} &:&\Delta \approx -21^{\circ },\eta \approx 0.48,\theta
\approx 31^{\circ }  \label{32}
\end{eqnarray}%
Hence we get the following bounds%
\begin{eqnarray}
\pi ^{+}D^{-}(\pi ^{-}D^{+}) &:&0\leq \delta _{f,\bar{f}}-\Delta \leq
26^{\circ }  \notag \\
\pi ^{-}K^{+}\text{ or }\pi ^{0}K^{0} &:&0\leq \delta _{f}-\Delta \leq
29^{\circ }  \label{33} \\
\pi ^{+}\pi ^{-} &:&0\leq \delta _{f}-\Delta \leq 31^{\circ }  \notag
\end{eqnarray}

Further we note that for these decays, $b$-quark is converted into $c$ or$u$
quark : $b\rightarrow c(u)+\bar{u}+d(s)$. In particular for the tree graph,
the configuration is such that $\bar{u}$ and $d(s)$ essentially go together
into a color singlet state with the third quark $c(u)$ recoiling; there is a
significant probability that the system will hadronize as a two body final
state \cite{9}. This physical picture has been put on the strong theoretical
basis \cite{10,11}, where in these references the QCD factorization have
been proved. For the tree amplitude, factorization implies $\delta
_{f}^{T}=0.$ We, therefore take the point of view that effective final state
phase shift is given by $\delta _{f}-\Delta .$ We take the lower bound for
the tree amplitude so that final state effective phase shift $\delta
_{f}^{T}=0.$ Thus for $\pi ^{+}D^{-}(\pi ^{-}D^{+}),\delta _{f}^{T}=\delta
_{f}^{\prime T}=0.$

The decay $B^{0}\rightarrow \pi ^{-}K^{+}$ is described by two amplitudes 
\cite{7}%
\begin{equation}
A(B^{0}\rightarrow \pi ^{-}K^{+})=-\left[ P+e^{i\gamma }T\right] =\left\vert
P\right\vert \left[ 1-re^{i(\gamma +\delta _{+-})}\right]  \label{16e}
\end{equation}%
where 
\begin{equation*}
P=-\left\vert P\right\vert e^{-i\delta _{P}},\text{ }T=\left\vert
T\right\vert e^{i\delta _{T}}\text{, }\delta _{+-}=\delta _{P}\text{, }r=%
\frac{\left\vert T\right\vert }{\left\vert P\right\vert }
\end{equation*}%
The decay $B^{0}\rightarrow \pi ^{0}K^{0}$ is described by the two
amplitudes \cite{7} 
\begin{equation}
A(B^{0}\rightarrow \pi ^{0}K^{0})=-\frac{1}{\sqrt{2}}\left\vert P\right\vert %
\left[ 1+r_{0}e^{i\left( \gamma +\delta _{00}\right) }\right]  \label{16f}
\end{equation}%
where 
\begin{equation*}
C=\left\vert C\right\vert e^{i\delta _{C}},\text{ }\delta _{00}=\delta
_{C}+\delta _{P},\text{ }r_{0}=\frac{\left\vert C\right\vert }{\left\vert
P\right\vert }
\end{equation*}%
For these decays, we use the lower bounds in Eq.(\ref{33}) for the tree
amplitude so that the effective final state phase $\delta _{T}=0.$ The phase 
$\delta _{C}$ is generated by rescattering correction and its value is -8$%
^{\circ }.$ For the direct $CP$ asymmetries, the relevant phases are $\delta
_{+-}$ and $\delta _{00}$. For the penguin amplitude, we assume that the
effective final state phase $\delta _{P}$ has the value near the upper
bound. Thus we have $\delta _{+-}\approx 29^{\circ },$ $\delta _{00}\approx
21^{\circ }.$

Now \cite{7}

\begin{eqnarray}
A_{CP}(B^{0} &\rightarrow &\pi ^{-}K^{+})=-\frac{2r\sin \gamma \sin \delta
_{+-}}{R}  \notag \\
R &=&1-2r\cos \gamma \cos \delta _{+-}+r_{+-}^{2}  \label{016g}
\end{eqnarray}%
Neglecting the terms of order $r^{2}$, we have 
\begin{equation}
\tan \gamma \tan \delta _{+-}=\frac{-A_{CP}(B^{0}\rightarrow \pi ^{-}K^{+})}{%
1-R}  \label{16h}
\end{equation}%
For $B^{0}\rightarrow \pi ^{0}K^{0}$%
\begin{eqnarray}
A_{CP}(B^{0} &\rightarrow &\pi ^{0}K^{0})=(R_{0}-1)\tan \gamma \tan \delta
_{00}  \label{16i} \\
R_{0} &=&1+2r_{0}\cos \gamma \cos \delta _{00}+r_{00}^{2}  \notag
\end{eqnarray}%
Now the experimental values of $A_{CP}$, $R$ and $R_{0}$ are \cite{12}%
\begin{eqnarray*}
A_{CP}(B^{0} &\rightarrow &\pi ^{-}K^{+})=-0.101\pm 0.015\text{ }(-0.097\pm
0.012) \\
A_{CP}(B^{0} &\rightarrow &\pi ^{0}K^{0})=-0.14\pm 0.11\text{ }(-0.00\pm \pm
0.10) \\
R &=&0.899\pm 0.048 \\
R_{0} &=&0.908\pm 0.068
\end{eqnarray*}%
where the numerical values in the bracket are the latest experimental values
as given in ref \cite{7}. With $\delta _{+-}\approx 29^{\circ },$ we get
from Eq.(\ref{16h}), $\gamma =(60\pm 3)^{\circ }.$ However for $\delta
_{+-}\approx 20^{\circ }$ which one gets from Eq.(\ref{15}) for $\rho
^{2}=0.65,\gamma =(69\pm 3)^{\circ }.$We obtain the following values for $%
A_{CP}(B^{0}\rightarrow \pi ^{0}K^{0})$ from Eqs.(\ref{16h}) and (\ref{16i})%
\begin{eqnarray*}
A_{CP}(B^{0} &\rightarrow &\pi ^{0}K^{0})=\frac{(1-R_{0})\tan \delta _{00}}{%
\left( 1-R\right) \tan \delta _{+-}}A_{CP}(B^{0}\rightarrow \pi ^{-}K^{+}) \\
&=&\left\{ 
\begin{array}{c}
\begin{array}{c}
-0.06\pm 0.01,\text{ \ \ }\delta _{+-}=29^{\circ } \\ 
\delta _{00}=21^{\circ } \\ 
-0.05\pm 0.01,\text{ \ \ }\delta _{+-}=20^{\circ }%
\end{array}
\\ 
\delta _{00}=12^{\circ }%
\end{array}%
\right\}
\end{eqnarray*}

We conclude: The phase shift $\delta _{+-}\approx (20-29)^{\circ }$ for $\pi
^{-}K^{+}$ is compatible with experimental value of the direct $CP-$%
asymmetry for $\pi ^{-}K^{+}$ decay mode. For $\pi ^{+}\pi ^{-},\delta
_{+-}\sim 31^{\circ }$ is compatible with the value ($33\pm 7_{-10}^{+8}$)$%
^{\circ }$ obtained by the authors of ref.[7]. Finally we note that the
actual value of the effective phase shift ($\delta _{f}-\Delta )$ depends on
one free parameter $\rho $, factorization implies $\delta _{f}^{T}=0$ i.e. $%
\delta _{f}-\Delta =0$ for the tree amplitude; for the penguin amplitude, $%
\delta _{f}^{P}$ depends on $\rho .$ However, from the experimental values
of the direct $CP$-violation for $\pi ^{-}K^{+},$ $\pi ^{-}\pi ^{+}$, it is
near the upper bound.

\ \ \ Finally we note that $\pi ^{+}D^{-}(\pi ^{-}D^{+}),\pi ^{-}K^{+},\pi
^{-}\pi ^{+}$ decays are $s$-wave decay whereas $B^{0}\rightarrow \rho
^{+}\pi ^{-}(\rho ^{-}\pi ^{+})$ decays are $p-$wave decays. For $p-$wave,
the decay amplitude 
\begin{eqnarray*}
f &=&\frac{1}{16\pi s}\int_{-s}^{0}M(s,t)(1+\frac{2t}{s})dt \\
&=&\frac{1}{16\pi s}iC_{P}\left[ \frac{1}{b}+\frac{2}{b^{2}}\frac{1}{s}%
\right] (s/s_{0}) \\
&&+\frac{2\gamma _{0}^{2}}{16\pi }i\left[ \frac{1}{\ln (s/s_{0})-i\pi }-%
\frac{2}{s}\frac{1}{\left[ \ln (s/s_{0})-i\pi \right] ^{2}}(s/s_{0})^{-1/2}%
\right] \\
&\approx &\frac{1}{16\pi s}iC_{P}\frac{1}{b}(s/s_{0})+\frac{2\gamma _{0}^{2}%
}{16\pi }i\frac{1}{\ln (s/s_{0})-i\pi }(s/s_{0})^{-1/2}+O\left( \frac{1}{s}%
\right)
\end{eqnarray*}%
to be compared with Eq.(\ref{31a}). Now for the $B\rightarrow \rho \pi $
decay, only longitudinal polarization of $\rho $ is effectively involved.
Since the longitudinal $\rho $-meson emulates a pseudoscalar meson and if we
assume same couplings as for pions, we conclude that the final state phase
for $\rho \pi $ should be of the order $30^{\circ }$; in any case it should
not be greater than $30^{\circ }$. The upper bound $\delta _{f}\leq 30^{0}$
can be used to select the several possible solutions in Table-2 [Section-4]
obtained from the analysis of weak decays $B\rightarrow \rho ^{+}\pi
^{-}\left( \rho ^{-}\pi ^{+}\right) $.

\section{CP Asymmetries and Strong Phases}

In this section, we discuss the experimental tests to verify the equality
(implied by C-invariance of S-matrix) of phase shifts $\delta _{f}$ and $%
\delta _{\bar{f}}$ for the weak decays of B mesons mentioned in section 1.%
\newline
It is convenient to write the time-dependent decay rates in the form \cite%
{13,6} 
\begin{eqnarray}
&&\left[ \Gamma _{f}(t)-\bar{\Gamma}_{\bar{f}}(t)\right] +\left[ \Gamma _{%
\bar{f}}-\bar{\Gamma}_{f}(t)\right]  \notag \\
&=&e^{-\Gamma t}\left\{ \cos \Delta mt\left[ \left( \left| A_{f}\right|
^{2}-\left| \bar{A}_{\bar{f}}\right| ^{2}\right) +\left( \left| A_{\bar{f}%
}\right| ^{2}-\left| \bar{A}_{f}\right| ^{2}\right) \right] \right.  \notag
\\
&&\left. +2\sin \Delta mt\left[ \text{Im}\left( e^{2i\phi _{M}}A_{f}^{\ast }%
\bar{A}_{f}\right) +\text{Im}\left( e^{2i\phi _{M}}A_{\bar{f}}^{\ast }\bar{A}%
_{\bar{f}}\right) \right] \right\}  \notag \\
&&  \label{e1} \\
&&\left[ \Gamma _{f}(t)+\bar{\Gamma}_{\bar{f}}(t)\right] -\left[ \Gamma _{%
\bar{f}}(t)+\bar{\Gamma}_{f}(t)\right]  \notag \\
&=&e^{-\Gamma t}\left\{ \cos \Delta mt\left[ \left( \left| A_{f}\right|
^{2}+\left| \bar{A}_{\bar{f}}\right| ^{2}\right) -\left( \left| A_{\bar{f}%
}\right| ^{2}+\left| \bar{A}_{f}\right| ^{2}\right) \right] \right.  \notag
\\
&&\left. +2\sin \Delta mt\left[ \text{Im}\left( e^{2i\phi _{M}}A_{f}^{\ast }%
\bar{A}_{f}\right) -\text{Im}\left( e^{2i\phi _{M}}A_{\bar{f}}^{\ast }\bar{A}%
_{\bar{f}}\right) \right] \right\}  \notag \\
&&  \label{e2}
\end{eqnarray}

\textbf{Case (i):} Eqs. $\eqref{e1}$ and $\eqref{e2}$ give 
\begin{eqnarray}
\mathcal{A}\left( t\right) &\equiv &\frac{[\Gamma _{f}(t)-\bar{\Gamma}_{\bar{%
f}}(t)]+[\Gamma _{\bar{f}}(t)-\bar{\Gamma}_{f}(t)]}{[\Gamma _{f}(t)+\bar{%
\Gamma}_{\bar{f}}(t)]+[\Gamma _{\bar{f}}(t)+\bar{\Gamma}_{f}]}  \notag \\
&=&\frac{2\bigl|F_{f}\bigr|\bigl|F_{\bar{f}}^{^{\prime }}\bigr|}{\bigl|F_{f}%
\bigr|^{2}+\bigl|F_{\bar{f}}^{^{\prime }}\bigr|^{2}}\sin \Delta mt\sin \bigl(%
2\phi _{M}-\phi -\phi ^{^{\prime }}\bigr)\cos \bigl(\delta _{f}-\delta _{%
\bar{f}}^{^{\prime }}\bigr)  \label{e6} \\
\mathcal{F}\left( t\right) &\equiv &\frac{\left[ \Gamma _{f}(t)+\bar{\Gamma}%
_{\bar{f}}\right] -\left[ \Gamma _{\bar{f}}(t)+\bar{\Gamma}_{f}\right] }{%
\left[ \Gamma _{f}(t)+\bar{\Gamma}_{\bar{f}}\right] +\left[ \Gamma _{\bar{f}%
}(t)+\bar{\Gamma}_{{f}}\right] }  \notag \\
&=&\frac{\bigl|F_{f}\bigr|^{2}-\bigl|F_{\bar{f}}^{^{\prime }}\bigr|^{2}}{%
\bigl|F_{f}\bigr|^{2}+\bigl|F_{\bar{f}}^{^{\prime }}\bigr|^{2}}\cos \Delta mt
\notag \\
&-&\frac{2\bigl|F_{f}\bigr|\bigl|F_{\bar{f}}^{^{\prime }}\bigr|}{\bigl|F_{f}%
\bigr|^{2}+\bigl|F_{\bar{f}}^{^{\prime }}\bigr|^{2}}\sin \Delta mt\cos
\left( 2\phi _{M}-\phi -\phi ^{^{\prime }}\right) \sin \bigl(\delta
_{f}-\delta _{\bar{f}}^{^{\prime }}\bigr)  \label{e7}
\end{eqnarray}

The effective Lagrangians $\mathcal{L}_{W}$ and $\mathcal{L}_{W}^{^{\prime}}$
are given by $(q = d, s)$ 
\begin{align}
\mathcal{L}_{W} &= V_{cb} V_{uq}^{\ast} [\bar{q} \gamma^{\mu} (1 -
\gamma^{5}) u] [\bar{c} \gamma_{\mu} (1 - \gamma_{5}) b]  \notag \\
\mathcal{L}_{W}^{^{\prime}} &= V_{ub} V_{cq}^{\ast} [\bar{q} \gamma^{\mu} (1
- \gamma^{5}) c] [\bar{u} \gamma_{\mu} (1 - \gamma_{5}) b]  \label{cc1} \\
\end{align}

Hence for these decays 
\begin{equation*}
\phi=0, \qquad \phi^{\prime}=\gamma
\end{equation*}

and 
\begin{equation}
\phi _{M}=%
\begin{cases}
-\beta, & \text{for $B^{0}$} \\ 
-\beta_{s}, & \text{for $B_{s}^{0}$}%
\end{cases}
\label{cc2}
\end{equation}
\begin{align}
A_{f}& =\langle D^{-}\pi ^{+}\left| \mathcal{L_{W}}\right| B^{0}\rangle
=F_{f}  \notag \\
\overset{^{\prime }}{A}_{\bar{f}}& =\langle D^{+}\pi ^{-}\left| \mathcal{%
L_{W}}^{\prime }\right| B^{0}\rangle =e^{i\gamma }\overset{^{\prime }}{F}_{%
\bar{f}}  \notag \\
A_{f_{s}}& =\langle K^{+}D_{s}^{-}\left| \mathcal{L_{W}}\right|
B_{s}^{0}\rangle =F_{f_{s}}  \notag \\
\overset{^{\prime }}{A}_{\bar{f}_{s}}& =\langle K^{-}D_{s}^{+}\left| 
\mathcal{L_{W}}^{\prime }\right| B_{s}^{0}\rangle =e^{i\gamma }\overset{%
^{\prime }}{F}_{\bar{f}_{s}}  \label{cc3}
\end{align}

Thus, we get from Eqs. $\eqref{e6} - \eqref{cc3}$ for $B^{0}$ decays, 
\begin{align}
\mathcal{A}\left(t\right)&=-\frac{2r_{D}}{1+r_{D}^{2}}\sin \Delta m_{B}t
\sin \left(2\beta+\gamma \right) \cos \left( \delta _{f}-\delta _{\bar{f}%
}^{^{\prime}}\right)  \notag \\
\mathcal{F}\left(t\right)&=\frac{1-r_{D}^{2}}{1+r_{D}^{2}}\cos \Delta m_{B}t
-\frac{2r_{D}}{1+r_{D}^{2}}\sin \Delta m_{B}t\cos \left(2\beta+\gamma
\right) \sin \left(\delta _{f}-\delta _{\bar{f}}^{^{\prime}}\right)
\label{4.35b}
\end{align}

\begin{equation}
\mathcal{A} = \frac{-2r_{D}}{1 + r_{D}^{2}} \sin (2\beta + \gamma) \frac{%
(\Delta m_{B}/\Gamma)}{1 + (\Delta m_{B}/\Gamma)^{2}} \cos (\delta_{f} -
\delta_{\bar{f}}^{^{\prime}})  \label{cc4}
\end{equation}

where 
\begin{equation}
r_{D} = \lambda^{2} R_{b} \frac{|F_{\bar{f}}^{^{\prime}}|}{|F_{f}|}
\label{cc5}
\end{equation}

For the decays, 
\begin{eqnarray*}
\bar{B}_{s}^{0}\left( B_{s}^{0}\right) &\rightarrow &D_{s}^{+} K^{-} \left(
D_{s}^{-} K^{+}\right) \\
\bar{B}_{s}^{0}\left( B_{s}^{0}\right) &\rightarrow &D_{s}^{-} K^{+} \left(
D_{s}^{+} K^{-}\right)
\end{eqnarray*}

we get, 
\begin{align}
\mathcal{A}_{s}\left( t\right) & =-\frac{2r_{D_{s}}}{1+r_{D_{s}}^{2}}\sin
\Delta m_{B_{s}}t\sin \left( 2\beta _{s}+\gamma \right) \cos \left( \delta
_{f_{s}}-\delta _{\bar{f}_{s}}^{^{\prime }}\right)  \notag \\
\mathcal{F}_{s}(t)& =\frac{1-r_{D_{s}}^{2}}{1+r_{D_{s}}^{2}}\cos \Delta
m_{B_{s}}t-\frac{2r_{D_{s}}}{1+r_{D_{s}}^{2}}\sin \Delta m_{B_{s}}t\cos
\left( 2\beta _{s}+\gamma \right) \sin \left( \delta _{f_{s}}-\delta _{\bar{f%
}_{s}}^{^{\prime }}\right)  \label{4.35}
\end{align}

where 
\begin{equation}
r_{D_{s}}=R_{b}\frac{|F_{\bar{f}_{s}}^{^{\prime }}|}{|F_{f_{s}}|}
\label{cc6}
\end{equation}

We note that for time integrated $CP$-asymmetry, 
\begin{align}
\mathcal{A}_{s}\equiv & \frac{\int_{0}^{\infty }\left[ \Gamma _{fs}\left(
t\right) -\bar{\Gamma}_{fs}\left( t\right) \right] dt}{\int_{0}^{\infty }%
\left[ \Gamma _{fs}\left( t\right) +\bar{\Gamma}_{fs}\left( t\right) \right]
dt}  \notag \\
=& -\frac{2r_{D_{s}}r}{1+r_{D_{s}}^{2}}\sin \left( 2\beta _{s}+\gamma
\right) \frac{\Delta m_{B_{s}}/\Gamma _{s}}{1+\left( \Delta m_{B_{s}}/\Gamma
_{s}\right) ^{2}}\cos (\delta _{f_{s}}-\delta _{\bar{f}_{s}}^{^{\prime }})
\label{4.35a}
\end{align}

The experimental results for the B decays are as follows \cite{12} 
\begin{equation}
\begin{array}{cccc}
& D^{-}\pi ^{+}\qquad & D^{\ast -}\pi ^{+}\qquad & D^{-}\rho ^{+} \\ 
\frac{S_{-}+S_{+}}{2}: & -0.046\pm 0.023 & -0.037\pm 0.012 & -0.024\pm
0.031\pm 0.009 \\ 
\frac{S_{-}-S_{+}}{2}: & -0.022\pm 0.021 & -0.006\pm 0.016 & -0.098\pm
0.055\pm 0.018%
\end{array}
\label{cc7}
\end{equation}

where 
\begin{align}
\frac{S_{-} + S_{+}}{2} \equiv & -\frac{2r_{D}}{1 + r^{2}_{D}} \sin(2\beta +
\gamma) \cos(\delta_{f} - \delta^{^{\prime}}_{\bar{f}})  \notag \\
\frac{S_{-} - S_{+}}{2} \equiv & -\frac{2r_{D}}{1 + r^{2}_{D}} \cos(2\beta +
\gamma) \sin(\delta_{f} - \delta^{^{\prime}}_{\bar{f}})  \label{cc8}
\end{align}

%

For $B_{s}^{0} \rightarrow D_{s}^{\ast -}K^{+}, D_{s}^{-}K^{+},
D_{s}^{-}K^{\ast +}$, replace $r_{D} \rightarrow r_{s}$, $\beta \rightarrow
\beta_{s}$, $\delta_{f} \rightarrow \delta_{f_{s}}$, $\delta^{^{\prime}}_{%
\bar{f}} \rightarrow \delta^{^{\prime}}_{\bar{f}_{s}}$ in Eq. $\eqref{cc8}$. 

Since for $B_{s}^{0}$, in the standard model, with three generations, gives $%
\beta_{s} = 0$, so we have for the CP-asymmetries $\sin\gamma$ or $%
\cos\gamma $ instead of $\sin(2\beta + \gamma)$, $\cos(2\beta + \gamma)$.
Hence $B_{s}^{0}$-decays are more suitable for testing the equality of phase
shifts $\delta_{f_{s}}$ and $\delta^{^{\prime}}_{\bar{f}_{s}}$ as for this
case neither $r_{s}$ nor $\cos\gamma$ is suppressed as compared to the
corresponding quantities for $B^{0}$. To conclude, for $B_{q}^{0}$ decays,
the equality of phases $\delta_{f}$ and $\delta^{^{\prime}}_{\bar{f}}$ for $%
B_{d}^{0}$ gives 
\begin{align}
-\frac{S_{-} + S_{+}}{2} &= 2r_{D} \sin(2\beta + \gamma)  \notag \\
-\frac{S_{-} - S_{+}}{2} &= 0  \label{cc9}
\end{align}

whereas for $B_{s}^{0}$ decays, we get 
\begin{align}
-\frac{S_{-}+S_{+}}{2}& =\frac{2r_{D_{s}}}{1+r_{D_{s}}^{2}}\sin (2\beta
_{s}+\gamma )  \notag \\
-\frac{S_{-}-S_{+}}{2}& =0  \label{cc10}
\end{align}

Corresponding to the decays $B_{s}^{0} \rightarrow D_{s}^{-}K^{+},
D_{s}^{+}K^{-}$ described by the tree diagrams, we have the color suppressed
decays $B^{0} \rightarrow \bar{D}^{0}K^{0}, D^{0}K^{0}$. For these decays, 
\begin{align*}
-\frac{S_{-} + S_{+}}{2} =& \frac{2r_{DK}}{1 + r_{DK}^{2}} \sin (2\beta +
\gamma) \cos(\delta_{\bar{D}^{0}K^{0}_{s}} - \delta^{^{\prime}}_{D^{0}\bar{K}%
^{0}_{s}}) \\
-\frac{S_{-} - S_{+}}{2} =& \frac{2r_{DK}}{1 + r_{DK}^{2}} \cos (2\beta +
\gamma) \sin(\delta_{\bar{D}^{0}K^{0}_{s}} - \delta^{^{\prime}}_{D^{0}\bar{K}%
^{0}_{s}}) \\
r_{DK} =& R_{b} \frac{\bigl|C_{D^{0}K_{s}}^{^{\prime}}\bigr|}{\bigl|C_{\bar{D%
}^{0}K_{s}}\bigr|}
\end{align*}

and the corresponding expression for $B_{s}^{0} \rightarrow \bar{D}^{0}\phi,
D^{0}\phi$. For the color suppressed decays $B^{0} \rightarrow \bar{D}%
^{0}\pi^{0}, D^{0}\pi^{0}$, we get similar expression as for $B^{0}
\rightarrow D^{-}\pi^{+}, D^{+}\pi^{-}$, with 
\begin{equation*}
r_{D} \equiv r_{D^{-}\pi^{-}}, \delta_{D^{-}\pi^{+}},
\delta^{^{\prime}}_{D^{-}\pi^{+}} \quad \text{replaced by} \quad
r_{D^{0}\pi^{0}}, \delta_{\bar{D}^{0}\pi^{0}},
\delta^{^{\prime}}_{D^{0}\pi^{0}}
\end{equation*}

To determine the parameter $r_{D}$ or $r_{D_{s}}$, we assume factorization
for the tree amplitude [7]. Factorization gives for the decays $\bar{B}%
^{0}\rightarrow D^{+}\pi ^{-},D^{\ast +}\pi ^{-},D^{+}\rho
^{-},D^{+}a_{1}^{-}$: 
\begin{align}
|\bar{F}_{\bar{f}}|=|\bar{T}_{\bar{f}}|& =G[f_{\pi
}(m_{B}^{2}-m_{D}^{2})f_{0}^{B-D}(m_{\pi }^{2}),2f_{\pi }m_{B}|\vec{p}%
|A_{0}^{B-D^{\ast }}(m_{\pi }^{2}),  \notag \\
& 2f_{\rho }m_{B}|\vec{p}|f_{+}^{B-D}(m_{\rho }^{2}),2f_{a_{1}}m_{B}|\vec{p}%
|f_{+}^{B-D}(a_{1}^{2})]  \label{c1} \\
|\bar{F}_{f}^{^{\prime }}|=|\bar{T}_{f}^{^{\prime }}|& =G^{^{\prime
}}[f_{D}(m_{B}^{2}-m_{\pi }^{2})f_{0}^{B-\pi }(m_{D}^{2}),2f_{D^{\ast
}}m_{B}|\vec{p}|f^{B-\pi }(m_{D^{\ast }}^{2}),  \notag \\
& 2f_{D}m_{B}|\vec{p}|A_{0}^{B-\rho }(m_{D}^{2}),2f_{D}m_{B}|\vec{p}%
|A_{0}^{B-a_{1}}(m_{B}^{2})]  \label{c2} \\
G& =\frac{G_{F}}{\sqrt{2}}|V_{ud}||V_{cb}|a_{1},\quad G^{^{\prime }}=\frac{%
G_{F}}{\sqrt{2}}|V_{cd}||V_{ub}|  \label{c3}
\end{align}

The decay widths for the above channels are given in the table 1 
\begin{table}[ht]
\centering                     
\begin{tabular}{|c|c|c|c|}
\hline
Decay & Decay Width $(10^{-9}$ MeV $\times |V_{cb}|^{2}$) & Form Factor & 
Form Factors $h(w^{(\ast)})$ \\ \hline
$\bar{B}^{0} \rightarrow D^{+} \pi^{-}$ & $(2.281) |f_{0}^{B-D}
(m_{\pi}^{2})|^{2}$ & $0.58 \pm 0.05$ & $0.51 \pm 0.03$ \\ \hline
$\bar{B}^{0} \rightarrow D^{\ast +} \pi^{-}$ & $(2.129) |A_{0}^{B-D^{*}}
(m_{\pi}^{2})|^{2}$ & $0.61 \pm 0.04$ & $0.54 \pm 0.03$ \\ \hline
$\bar{B}^{0} \rightarrow D^{+} \rho^{-}$ & $(5.276) |f_{+}^{B-D}
(m_{\rho}^{2})|^{2}$ & $0.65 \pm 0.11 $ & $0.57 \pm 0.10$ \\ \hline
$\bar{B}^{0} \rightarrow D^{+} a_{1}^{-}$ & $(5.414) |f_{+}^{B-D}
(m_{a_{1}}^{2})|^{2}$ & $0.57 \pm 0.31 $ & $0.50 \pm 0.27$ \\ \hline
\end{tabular}%
\caption{Form Factors}
\label{tab:1}
\end{table}

where we have used 
\begin{equation*}
a_{1}^{2} |V_{ud}|^{2} \approx 1, \quad f_{\pi} = 131 MeV, \quad f_{\rho} =
209 MeV, \quad f_{a_{1}} = 229 MeV
\end{equation*}

Using the experimental branching ratios and \cite{12} 
\begin{equation}
|V_{cb}|=(38.3\pm 1.3)\times 10^{-3}  \label{c5}
\end{equation}

we obtain the corresponding form factors given in Table 1.

In terms of variables \cite{14,15}: 
\begin{equation}
\omega =v\cdot v^{^{\prime }},\quad v^{2}=v^{^{\prime }2}=1,\quad
t=q^{2}=m_{B}^{2}+m_{D^{(\ast )}}^{2}-2m_{B}m_{D^{(\ast )}}\omega  \label{c7}
\end{equation}

the form factors can be put in the following form 
\begin{align}
f_{+}^{B-D}(t) &= \frac{m_{B} + m_{D}}{2\sqrt{m_{B}m_{D}}} h_{+}(\omega),
\quad f_{0}^{B-D}(t) = \frac{\sqrt{m_{B}m_{D}}}{m_{B} + m_{D}} (1 +
\omega)h_{0}(\omega)  \notag \\
A_{2}^{B-D^{\ast}} (t) &= \frac{m_{B} + m_{D^{\ast}}}{2\sqrt{%
m_{B}m_{D^{\ast}}}} (1 + \omega) h_{A_{2}}(\omega), \quad A_{0}^{B-D^{\ast}}
(t) = \frac{m_{B} + m_{D^{\ast}}}{2\sqrt{m_{B}m_{D^{\ast}}}} h_{A_{0}}
(\omega)  \notag \\
A_{1}^{B-D^{\ast}} (t) &= \frac{\sqrt{m_{B}m_{D^{\ast}}}}{m_{B} +
m_{D^{\ast}}} (1 + \omega) h_{A_{1}} (\omega)  \label{c8}
\end{align}

Heavy Quark Effective Theory (HQET) gives \cite{14,15}: 
\begin{equation*}
h_{+}(\omega )=h_{0}(\omega )=h_{A_{0}}(\omega )=h_{A_{1}}(\omega
)=h_{A_{2}}(\omega )=\zeta (\omega )
\end{equation*}

where $\zeta (\omega )$ is the form factor, with normalization $\zeta (1)=1$%
. For 
\begin{align}
t& =m_{\pi }^{2},m_{\rho }^{2},m_{a_{1}}^{2}  \notag \\
\omega ^{(*)}& =1.589(1.504),1.559,1.508  \label{ccc1}
\end{align}

In reference \cite{16}, the value quoted for $h_{A_{1}}(\omega _{max}^{\ast
})$ is 
\begin{equation}
|h_{A_{1}}(\omega _{max}^{\ast })|=0.52\pm 0.03  \label{c10}
\end{equation}

Since $\omega _{max}^{*}=1.504$, the value for $|h_{A_{0}}(\omega _{\max
}^{*})|$ obtained in Table 1 is in remarkable agreement with the value given
in Eq. $\eqref{c10}$ showing that factorization assumption for $%
B^{0}\rightarrow \pi D^{(*)}$ decays is experimentally on solid footing and
is in agreement with HQET.

From Eqs. $\eqref{c1}$ and $\eqref{c2}$, we obtain 
\begin{align}
r_{D}& =\lambda ^{2}R_{b}\frac{|\bar{T}_{f}^{^{\prime }}|}{|\bar{T}_{\bar{f}%
}|}  \notag \\
& =\lambda ^{2}R_{b}\left[ \frac{f_{D}(m_{B}^{2}-m_{\pi }^{2})f_{0}^{B-\pi
}(m_{D}^{2})}{f_{\pi }(m_{B}^{2}-m_{D}^{2})f_{0}^{B-D}(m_{\pi }^{2})},\quad 
\frac{f_{D^{\ast }}f_{+}^{B-\pi }(m_{D^{\ast }}^{2})}{f_{\pi
}A_{0}^{B-D}(m_{\pi }^{2})},\quad \frac{f_{D}A_{0}^{B-\rho }(m_{D}^{2})}{%
f_{\rho }f_{+}^{B-D}(m_{\rho ^{2}})}\right]  \label{c11}
\end{align}

where 
\begin{equation}
\frac{|V_{ub}| |V_{cd}|}{|V_{cb}| |V_{ud}|} = \lambda^{2} R_{b} \approx
(0.227)^{2} (0.40) \approx 0.021  \label{c12}
\end{equation}

To determine $r_{D}$, we need information for the form factors $f_{0}^{B-\pi
}(m_{D}^{2}),f_{+}^{B-\pi }(m_{D}^{2}),A_{0}^{B-\rho }(m_{D}^{2})$. For
these form factors, we use the following values \cite{17,18}: 
\begin{align*}
A_{0}^{B-\rho }(0)& =0.30\pm 0.03,A_{0}^{B-\rho }(m_{D}^{2})=0.38\pm 0.04 \\
f_{+}^{B-\pi }(0)& =f_{0}^{B-\pi }(0)=0.26\pm 0.04,\quad f_{+}^{B-\pi
}(m_{D^{\ast }}^{2})=0.32\pm 0.05,\quad f_{0}^{B-D}(m_{D}^{2})=0.28\pm 0.04
\\
&
\end{align*}

Along with the values of remaining form factors given in Table 1, we obtain 
\begin{equation}
r_{D^{(\ast )}}=[0.018\pm 0.002,\quad 0.017\pm 0.003,\quad 0.012\pm 0.002]
\label{c13}
\end{equation}

The above value for $r_{D}^{\ast}$ gives 
\begin{equation}
- \left( \frac{S_{+} + S_{-}}{2} \right)_{D^{\ast} \pi} = 2 (0.017 \pm
0.003) \sin (2\beta + \gamma)  \label{c15}
\end{equation}

The experimental value of the CP asymmetry for $B^{0} \rightarrow D^{\ast}
\pi$ decay has the least error. Hence we obtain the following bounds 
\begin{align}
\sin (2\beta + \gamma) &> 0.69  \label{c16} \\
44^{\circ} &\leq (2\beta + \gamma) \leq 90^{\circ} \\
\text{or} \quad 90^{\circ} &\leq (2\beta + \gamma) \leq 136^{\circ}
\end{align}

Selecting the second solution, and using $2\beta \approx 43^{\circ }$, we
get 
\begin{equation}
\gamma =(70\pm 23)^{\circ }  \label{c19}
\end{equation}

Further, we note that the factorization for the decay $\bar{B}^{0}
\rightarrow D_{s}^{\ast -} \pi^{+}$ gives 
\begin{equation}
\bar{T} = |V_{ub}| |V_{cs}| f_{D_{s}^{\ast}} 2 m_{B} |\vec{p}| f_{+}^{B-\pi}
(m_{D_{s}^{\ast}}^{2})  \label{c20}
\end{equation}

Using the experimental branching ratio for this decay, we get 
\begin{equation}
\left( \frac{f_{D_{s}^{\ast}}}{f_{\pi}} \right)^{2} \left| \frac{%
f_{+}^{B-\pi}(m_{D_{s}^{\ast}}^{2})}{f_{+}^{B-\pi}(0)} \right|^{2} = 7.7 \pm
1.9  \label{c21}
\end{equation}

On using 
\begin{equation}
\frac {f_{+}^{B-\pi}(0)} {f_{+}^{B-\pi} (m_{D_{s}^{\ast}}^{2})} = 0.77 \pm
0.09  \label{c22}
\end{equation}

we get 
\begin{equation}
f_{D_{s}^{\ast}} = 279 \pm 79 MeV  \label{c23}
\end{equation}

Similar analysis for $\bar{B}^{0}\rightarrow D_{s}^{-}\pi ^{+}$ gives 
\begin{equation}
\left( \frac{f_{D_{s}}}{f_{\pi }}\right) ^{2}\left| \frac{f_{0}^{B-\pi
}(m_{D_{s}}^{2})}{f_{0}^{B-\pi }(0)}\right| ^{2}=2.72\pm 0.64  \label{c24}
\end{equation}

On using 
\begin{equation}
\frac{f_{0}^{B-\pi} (0)} {f_{0}^{B-\pi}(m_{D_{s}^{2}})} = 0.93 \pm 0.05
\label{c25}
\end{equation}

we get 
\begin{equation}
f_{D_{s}} = 201 \pm 47 MeV  \label{c26}
\end{equation}

Finally from the experimental branching ratio for the decay $\bar{B}_{s}^{0}
\rightarrow D_{s}^{+} \pi^{-}$, we obtain 
\begin{align}
f_{0}^{B_{s}-D_{s}} (0) &= 0.62 \pm 0.18  \label{c27} \\
h_{0} (1.531) &= 0.55 \pm 0.16  \label{c28}
\end{align}

To end this section, we discuss the decays $\bar{B}_{s}^{0}\rightarrow
D_{s}^{+}K^{-},D_{s}^{\ast +}K^{-}$ for which no experimental data are
available. However, using factorization, we get 
\begin{align}
\Gamma (\bar{B}_{s}^{0}\rightarrow D_{s}^{+}K^{-})& =(1.75\times
10^{-10})|V_{cb}f_{0}^{B_{s}-D_{s}}(m_{K}^{2})|^{2}MeV  \label{c31} \\
\Gamma (\bar{B}_{s}^{0}\rightarrow D_{s}^{\ast +}K^{-})& =(1.57\times
10^{-10})|V_{cb}A_{0}^{B_{s}-D_{s}^{\ast }}(m_{K}^{2})|^{2}MeV  \label{c32}
\end{align}

SU(3) gives 
\begin{align}
|V_{cb}f_{0}^{B_{s}-D_{s}}(m_{K}^{2})|^{2}& \approx
|V_{cb}||f_{0}^{B-D}(m_{\pi }^{2})|^{2}=(0.50\pm 0.04)\times 10^{-3}  \notag
\\
|V_{cb}A_{0}^{B_{s}-D_{s}^{\ast }}(m_{K}^{2})|^{2}& \approx
|V_{cb}||A_{0}^{B-D^{\ast }}(m_{\pi }^{2})|^{2}=(0.56\pm 0.04)\times 10^{-3}
\label{c33}
\end{align}

From the above equations, we get the following branching ratios 
\begin{equation}
\frac{\Gamma (\bar{B_{s}}^{0} \rightarrow D_{s}^{(\ast) + }K^{-})} {\Gamma_{%
\bar{B}_{s}^{0}}} = (1.94 \pm 0.07) \times 10^{-4} [(1.96 \pm 0.07) \times
10^{-4}]  \label{c34}
\end{equation}

For $\bar{B}_{s}^{0}\rightarrow D_{s}^{*+}K^{-}$ 
\begin{equation}
r_{D_{s}}=R_{b}\left[ \frac{f_{D_{s}^{*}}f_{+}^{B_{s}-K}(m_{D_{s}^{*}}^{2})}{%
f_{K}A_{0}^{B_{s}-D_{s}^{*}}(m_{K}^{2})}\right]  \label{c35}
\end{equation}

Hence we get 
\begin{align}
-(\frac{S_{+} + S_{-}}{2})_{D_{s}^{\ast} K} &= (0.41 \pm 0.08) \sin
(2\beta_{s} + \gamma)  \notag \\
&= (0.41 \pm 0.08) \sin \gamma  \label{37}
\end{align}

where we have used 
\begin{align}
R_{b} &= 0.40, \quad \frac{f_{D_{s}}}{f_{K}} = \frac{f_{D_{s}^{\ast}}}{f_{K}}
= 1.75 \pm 0.06, \quad f_{+}^{B_{s} - K}(m_{D_{s}^{\ast}}^{2}) = 0.34 \pm
0.06  \notag \\
A_{0}^{B_{s} - D_{s}^{\ast}} (m_{K}^{2}) &= A_{0}^{B_{s} - D_{s}^{\ast}} (0)
= \frac{m_{B_{s}} + m_{D_{s}^{\ast}}} {2\sqrt{m_{B_{s} m_{D_{s}^{\ast}}}}} %
\left[h_{0} (\omega_{s}^{\ast} = 1.453) = 0.52 \pm .03 \right]  \notag \\
&= 0.58 \pm 0.03  \label{c36}
\end{align}

\section{CP Asymmetries for $A_{f} \neq A_{\bar{f}}$}

We now discuss the decays listed in case (ii) where $A_{f} \neq A_{\bar{f}}$%
. Subtracting and adding Eqs. $(\ref{e2})$ and $(\ref{e1})$, we get, 
\begin{align}
\frac{\Gamma _{f}(t)-\bar{\Gamma}_{f}(t)}{\Gamma _{f}(t)+\bar{\Gamma}_{f}(t)}
=& C_{f}\cos\Delta mt + S_{f}\sin\Delta mt  \notag \\
=& (C-\Delta C) \cos \Delta mt + (S-\Delta S) \sin\Delta mt  \label{ccc2} \\
\frac{\Gamma _{\bar{f}}(t)-\bar{\Gamma}_{\bar{f}}(t)}{\Gamma _{\bar{f}}(t)+ 
\bar{\Gamma}_{\bar{f}}(t)} =& C_{\bar{f}}\cos\Delta mt + S_{\bar{f}%
}\sin\Delta mt  \notag \\
=& (C+\Delta C) \cos \Delta mt + (S+\Delta S) \sin \Delta mt  \label{ccc3}
\end{align}

where 
\begin{align}
C_{\bar{f},f} &= (C\pm\Delta C)  \notag \\
&= \frac{\bigl|A_{\bar{f},f}\bigr|^{2} - \bigl|\bar{A}_{\bar{f},f}\bigr|^{2}%
}{\bigl|A_{\bar{f},f}\bigr|^{2} + \bigl|\bar{A}_{\bar{f},f}\bigr|^{2}} 
\notag \\
&= \frac{\Gamma_{\bar{f},f}-\bar{\Gamma}_{\bar{f},f}}{\Gamma_{\bar{f},f} + 
\bar{\Gamma}_{\bar{f},f}}  \notag \\
&= \frac{R_{\bar{f}, f} (1 - A_{CP}^{\bar{f}, f}) - R_{\bar{f}, f} (1 +
A_{CP}^{\bar{f}, f})}{\Gamma (1 \pm A_{CP})}  \label{ccc4} \\
S_{\bar{f},f} &= (S \pm \Delta S) \\
&= \frac{2\text{Im} [e^{2i\phi_{M}}A^{\ast}_{\bar{f},f}\bar{A}_{\bar{f},f}]}{%
\Gamma_{\bar{f},f} + \bar{\Gamma}_{\bar{f},f}}  \label{ccc5} \\
A_{CP}^{\bar{f}} &= \frac{\bar{\Gamma}_{f}-\Gamma _{\bar{f}}}{\Gamma _{\bar{f%
}}+\bar{\Gamma}_{f}}  \notag \\
A_{CP}^{f} &= \frac{\bar{\Gamma}_{\bar{f}} - \Gamma _{f}}{\Gamma _{f} + \bar{%
\Gamma}_{\bar{f}}}  \label{ccc6} \\
A_{CP}&= \frac{(\Gamma_{\bar{f}} + \bar{\Gamma}_{\bar{f}}) - (\bar{\Gamma_{f}%
} + \Gamma_{f})}{(\Gamma_{\bar{f}} - \bar{\Gamma}_{\bar{f}}) - (\bar{%
\Gamma_{f}} + \Gamma_{f})} \\
&= \frac{R_{f}A^{f}_{CP}-R_{\bar{f}}A^{\bar{f}}_{CP}}{\Gamma}  \label{ccc7}
\end{align}

where 
\begin{align}
R_{f} &= \frac{1}{2}(\Gamma_{f} + \bar{\Gamma}_{\bar{f}}), \qquad R_{\bar{f}%
}=\frac{1}{2}(\Gamma_{\bar{f}} + \bar{\Gamma}_{f})  \notag \\
\Gamma &= R_{f}+R_{\bar{f}}  \label{ccc8}
\end{align}

The following relations are also useful which can be easily derived from
above equations 
\begin{align}
\frac{R_{\bar{f},f}}{R_{f}+R_{\bar{f}}}& =\frac{1}{2}[(1\pm \Delta C)\pm
A_{CP}C]  \label{ccc9} \\
\frac{R_{\bar{f}}-R_{f}}{R_{f}+R_{\bar{f}}}& =[\Delta C+A_{CP}C]
\label{ccc10} \\
\frac{R_{\bar{f}}A_{CP}^{\bar{f}}+R_{f}A_{CP}^{f}}{R_{f}+R_{\bar{f}}}&
=[C+A_{CP}\Delta C]  \label{ccc11}
\end{align}

For these decays, the decay amplitudes can be written in terms of tree
amplitude $e^{i\phi_{T}}T_{f}$ and the penguin amplitude $e^{i\phi_{P}}P_{f}$%
: 
\begin{align}
A_{f} &= e^{i\phi_{T}}e^{i\delta_{f}^{T}}\bigl|T_{f}\bigr| \lbrack 1 +
r_{f}e^{i(\phi_{P}-\phi_{T})}e^{i\delta_{f}}]  \notag \\
A_{\bar{f}} &= e^{i\phi_{T}}e^{i\delta_{\bar{f}}^{T}}\bigl|T_{\bar{f}}\bigr| %
\lbrack 1 + r_{\bar{f}} e^{i(\phi_{P}-\phi_{T})}e^{i\delta_{\bar{f}}}]
\label{ccc12}
\end{align}

where $r_{f, \bar{f}} = \frac{\bigl|P_{f,\bar{f}}\bigr|}{\bigl|T_{f,\bar{f}}%
\bigr|}, \quad \delta_{f, \bar{f}} = \delta^{P}_{f, \bar{f}} - \delta^{T}_{f,%
\bar{f}} $. 
\begin{align}
\bar{A}_{\bar{f}} &= e^{-i\phi_{T}}e^{i\delta_{f}^{T}}\bigl|T_{f}\bigr| %
\lbrack 1 + r_{f}e^{-i(\phi_{P}-\phi_{T})}e^{i\delta_{f}}]  \notag \\
\bar{A}_{f} &= e^{-i\phi_{T}}e^{i\delta_{\bar{f}}^{T}}\bigl|T_{\bar{f}}%
\bigr| \lbrack 1 + r_{\bar{f}} e^{-i(\phi_{P}-\phi_{T})}e^{i\delta_{\bar{f}%
}}]  \label{ccc13}
\end{align}

\begin{equation}
\text{For} B^{0} \rightarrow \rho^{-} \pi^{+}: A_{f}; \qquad B^{0}
\rightarrow \rho^{+} \pi^{-}: A_{\bar{f}}; \quad \phi_{T} = \gamma, \phi_{P}
= -\beta  \label{ccc14}
\end{equation}

\begin{equation}
\text{For} B^{0} \rightarrow D^{\ast -}D^{+}: A^{D}_{f}; \qquad B^{0}
\rightarrow D^{\ast +}D^{-}: A^{D}_{\bar{f}}; \quad \phi_{T} = 0, \phi_{P} =
-\beta  \label{ccc15}
\end{equation}

Hence for $B^{0}\rightarrow \rho ^{-}\pi ^{+},B^{0}\rightarrow \rho ^{+}\pi
^{-}$, we have 
\begin{align}
A_{f}& =\bigl|T_{f}\bigr|e^{+i\gamma }e^{i\delta
_{f}^{T}}[1-r_{f}e^{i(\alpha +\delta _{f})}]  \notag \\
A_{\bar{f}}& =\bigl|T_{\bar{f}}\bigr|e^{+i\gamma }e^{i\delta _{\bar{f}%
}^{T}}[1-r_{\bar{f}}e^{i(\alpha +\delta _{\bar{f}})}]  \label{ccc16} \\
\text{where}\qquad r_{f,\bar{f}}& =\frac{|V_{tb}||V_{td}|}{|V_{ub}||V_{ud}|}%
\frac{\bigl|P_{f,\bar{f}}\bigr|}{\bigl|T_{f,\bar{f}}\bigr|}=\frac{R_{t}}{%
R_{b}}\frac{\bigl|P_{f,\bar{f}}\bigr|}{\bigl|T_{f,\bar{f}}\bigr|}
\label{ccc17}
\end{align}

and for $\text{B}^{0}\rightarrow D^{*-}D^{+}$, $\text{B}^{0}\rightarrow
D^{*+}D^{-}$, we have 
\begin{align}
A_{f}^{D}& =\bigl|T_{f}^{D}\bigr| e^{i\delta
_{f}^{TD}}[1-r_{f}^{D}e^{i(-\beta +\delta _{f}^{D})}]  \notag \\
A_{\bar{f}}^{D}& =\bigl|T_{\bar{f}}^{D}\bigr| e^{i\delta _{\bar{f}%
}^{TD}}[1-r_{\bar{f}}^{D}e^{i(-\beta +\delta _{\bar{f}}^{D})}]  \label{ccc18}
\\
\text{where}\qquad r_{f,\bar{f}}& =R_{t}\frac{\bigl|P_{f,\bar{f}}^{D}\bigr|}{%
\bigl|T_{f,\bar{f}}^{D}\bigr|}  \notag
\end{align}

We now confine ourselves to $B^{0}(\bar{B}^{0})\rightarrow \rho ^{-}\pi
^{+},\rho ^{+}\pi ^{-}(\rho ^{+}\pi ^{-},\rho ^{-},\pi ^{+})$ decays only 
\cite{19,20}. The experimental results for these decays are \cite{12} as 
\begin{align}
\Gamma & =R_{f}+R_{\bar{f}}=(22.8\pm 2.5)\times 10^{-6}  \label{ccc19} \\
A_{CP}^{f}& =-0.16\pm 0.23,\quad A_{CP}^{\bar{f}}=0.08\pm 0.12  \label{ccc20}
\\
C& =0.01\pm 0.14,\quad \Delta C=0.37\pm 0.08  \label{ccc21} \\
S& =0.01\pm 0.09,\quad \Delta S=-0.05\pm 0.10  \label{ccc22}
\end{align}

With the above values, it is hard to draw any reliable conclusion.
Neglecting the term $A_{CP}C$ in Eqs. $\eqref{ccc9}$ and $\eqref{ccc10}$, we
get 
\begin{align}
R_{\bar{f},f}& =\frac{1}{2}\Gamma (1\pm \Delta C)  \label{ccc23} \\
R_{\bar{f}}-R_{f}& =\Delta C  \notag
\end{align}

Using the above value for $\Delta C$, we obtain 
\begin{align}
R_{\bar{f}} &= (15.6 \pm 1.7) \times 10^{-6}  \notag \\
R_{f} &= (7.2 \pm 0.8) \times 10^{-6}  \label{ccc25}
\end{align}

We analyze these decays by assuming factorization for the tree graphs \cite%
{10,11}. This assumption gives 
\begin{align}
T_{\bar{f}}& =\bar{T}_{f}\sim 2m_{B}f_{\rho }|\vec{p}|f_{+}(m_{\rho }^{2})
\label{ccc26} \\
T_{f}& =\bar{T}_{\bar{f}}\sim 2m_{B}f_{\pi }|\vec{p}|A_{0}(m_{\pi }^{2})
\label{ccc27}
\end{align}

Using $f_{+}(m_{\rho }^{2})\approx 0.26\pm 0.04$ and $A_{0}(m_{\pi
}^{2})\approx A_{0}(0)=0.29\pm 0.03$ and $|V_{ub}|=(3.5\pm 0.6)\times
10^{-3} $, we get the following values for the tree amplitude contribution
to the branching ratios 
\begin{align}
\Gamma _{\bar{f}}^{\text{tree}}& =(15.6\pm 1.1)\times 10^{-6}\equiv |T_{\bar{%
f}}|^{2}  \label{ccc28} \\
\Gamma _{f}^{\text{tree}}& =(7.6\pm 1.4)\times 10^{-6}\equiv |T_{f}|^{2}
\label{ccc29} \\
t& =\frac{T_{f}}{T_{\bar{f}}}=\frac{f_{\pi }A_{0}(m_{\pi }^{2})}{f_{\rho
}f_{+}(m_{\rho }^{2})}=0.70\pm 0.12  \label{ccc30}
\end{align}

Now 
\begin{align}
B_{\bar{f}} &= \frac{R_{\bar{f}}}{|T_{\bar{f}}|^{2}} = 1 - 2 r_{\bar{f}}
\cos \alpha \cos \delta_{\bar{f}} + r_{\bar{f}}^{2}  \label{ccc31} \\
B_{f} &= \frac{R_{f}}{|T_{f}|^{2}} = 1 - 2 r_{f} \cos \alpha \cos \delta_{f}
+ r_{f}^{2}  \label{ccc32}
\end{align}

Hence from Eqs. $\eqref{ccc25}$ and $\eqref{ccc29}$, we get 
\begin{align}
B_{\bar{f}} &= 1.00 \pm 0.12  \notag \\
B_{f} &= 0.95 \pm 0.11  \label{ccc33}
\end{align}

In order to take into account the contribution of penguin diagram, we
introduce the angles $\alpha _{eff}^{f,\bar{f}}$ \cite{21}, defined as
follows 
\begin{align}
e^{i\beta }A_{f,\bar{f}}& =|A_{f,\bar{f}}|e^{-i\alpha _{eff}^{f,\bar{f}}} 
\notag \\
e^{-i\beta }\bar{A}_{\bar{f},f}& =|\bar{A}_{\bar{f},f}|e^{i\alpha _{eff}^{f,%
\bar{f}}}  \label{ccc34}
\end{align}

With this definition, we separate out tree and penguin contributions: 
\begin{align}
e^{i\beta }A_{f,\bar{f}}-e^{-i\beta }\bar{A}_{\bar{f},f}& =|A_{f,\bar{f}%
}|e^{-i\alpha ^{f,\bar{f}}}-|\bar{A}_{\bar{f},f}|e^{i\alpha ^{f,\bar{f}}} 
\notag \\
& =2iT_{f,\bar{f}}\sin \alpha   \label{ccc35} \\
e^{i(\alpha +\beta )}A_{f,\bar{f}}-e^{-i(\alpha +\beta )}\bar{A}_{\bar{f}%
,f}& =|A_{f,\bar{f}}|e^{-i(\alpha _{eff}^{f,\bar{f}}-\alpha )}-|\bar{A}_{%
\bar{f},f}|e^{i(\alpha _{eff}^{f,\bar{f}}-\alpha )}  \notag \\
& =(2iT_{f,\bar{f}}\sin \alpha )r_{f,\bar{f}}e^{i\delta _{f,\bar{f}}}  \notag
\\
& =2iP_{f,\bar{f}}\sin \alpha   \label{ccc36}
\end{align}

From Eq. $\eqref{ccc35}$, we get 
\begin{align}
2\frac{|T_{f,\bar{f}}|^{2}}{R_{f,\bar{f}}}\sin ^{2}\alpha & \equiv \frac{%
2\sin ^{2}\alpha }{B_{f,\bar{f}}}=1-\sqrt{1-A_{CP}^{f,\bar{f}2}}\cos 2\alpha
_{eff}^{f,\bar{f}}  \label{ccc37} \\
\sin 2\delta _{f,\bar{f}}^{T}& =-A_{CP}^{f,\bar{f}}\frac{\sin 2\alpha
_{eff}^{f,\bar{f}}}{1-\sqrt{1-A_{CP}^{f,\bar{f}2}}\cos 2\alpha _{eff}^{f,%
\bar{f}}}  \label{ccc38a} \\
\cos 2\delta _{f,\bar{f}}^{T}& =\frac{\sqrt{1-A_{CP}^{f,\bar{f}2}}-\cos
2\alpha _{eff}^{f,\bar{f}}}{1-\sqrt{1-A_{CP}^{f,\bar{f}2}}\cos 2\alpha
_{eff}^{f,\bar{f}}}  \label{ccc38b}
\end{align}

From Eqs. $\eqref{ccc35}$ and $\eqref{ccc36}$, we get 
\begin{align}
r_{f,\bar{f}}^{2}& =\frac{1-\sqrt{1-A_{CP}^{f,\bar{f}2}}\cos (2\alpha
_{eff}^{f,\bar{f}}-2\alpha )}{1-\sqrt{1-A_{CP}^{f,\bar{f}2}}\cos 2\alpha
_{eff}^{f,\bar{f}}}  \label{ccc39} \\
r_{f,\bar{f}}\cos \delta _{f,\bar{f}}& =\frac{\cos \alpha -\sqrt{1-A_{CP}^{f,%
\bar{f}2}}\cos (2\alpha _{eff}^{f,\bar{f}}-\alpha )}{1-\sqrt{1-A_{CP}^{f,%
\bar{f}2}}\cos 2\alpha _{eff}^{f,\bar{f}}}  \label{ccc40} \\
r_{f,\bar{f}}\sin \delta _{f,\bar{f}}& =\frac{-\frac{A_{CP}^{f,\bar{f}}}{%
\sin \alpha }}{1-\sqrt{1-A_{CP}^{f,\bar{f}2}}\cos 2\alpha _{eff}^{f,\bar{f}}}
\label{ccc41}
\end{align}

Now factorization implies \cite{22} 
\begin{equation}
\delta _{f}^{T}=0=\delta _{\bar{f}}^{T}  \label{ccc42}
\end{equation}

Thus in the limit $\delta_{f}^{T} \rightarrow 0$, we get for Eq. $%
\eqref{ccc38b}$ 
\begin{align}
\cos 2 \alpha_{eff}^{f, \bar{f}} &= -1, \qquad \alpha_{eff}^{f, \bar{f}} =
90^{\circ}  \label{ccc43} \\
r_{f, \bar{f}} \cos \delta_{f, \bar{f}} &= \cos \alpha  \label{ccc44} \\
r_{f, \bar{f}} \sin \delta_{f, \bar{f}} &= \frac{-A_{CP}^{f, \bar{f}} / \sin
\alpha}{1 + \sqrt{1 - A_{CP}^{f, \bar{f} 2}}}  \label{ccc45} \\
r_{f, \bar{f}}^{2} &= \frac{1 + \sqrt{1 - A_{CP}^{f, \bar{f} 2}} \cos 2
\alpha}{1 + \sqrt{1 - A_{CP}^{f, \bar{f} 2}}}  \label{ccc47} \\
&\approx \cos^{2} \alpha + \frac{1}{4} A_{CP}^{f, \bar{f} 2} \sin^{2} \alpha
\label{ccc48}
\end{align}


The solution of Eq. $\eqref{ccc44}$ is graphically shown in Fig. 1 for $%
\alpha$ in the range $80^{\circ} \leq \alpha < 103^{\circ}$ for $r_{f, \bar{f%
}} = 0.10, 015, 0.20, 0.25, 0.30$. From the figure, the final state phases $%
\delta_{f, \bar{f}}$ for various values of $r_{f, \bar{f}}$ can be read for
each value of $\alpha$ in the above range. Few examples are given in Table 2

\begin{table}[ht]
\centering                     
\begin{tabular}{|c|c|c|c|}
\hline
$\alpha$ & $r_{f}$ & $\delta_{f}$ & $A_{CP}^{f} \approx -2 r_{f} \sin
\delta_{f} \sin \alpha$ \\ \hline
$80^{\circ}$ & 0.20 & $29^{\circ}$ & -0.19 \\ \hline
& 0.25 & $46^{\circ}$ & -0.36 \\ \hline
$82^{\circ}$ & 0.15 & $22^{\circ}$ & -0.11 \\ \hline
& 0.20 & $46^{\circ}$ & -0.28 \\ \hline
$85^{\circ}$ & 0.10 & $29^{\circ}$ & -0.10 \\ \hline
& 0.15 & $54^{\circ}$ & -0.24 \\ \hline
$86^{\circ}$ & 0.10 & $46^{\circ}$ & -0.14 \\ \hline
& 0.15 & $62^{\circ}$ & -0.26 \\ \hline
$88^{\circ}$ & 0.10 & $70^{\circ}$ & -0.19 \\ \hline
\end{tabular}%
\caption{}
\label{tab:2}
\end{table}

For $\alpha >90^{\circ }$, change $\alpha \rightarrow \pi -\alpha $, $\delta
_{f}\rightarrow \pi -\delta _{f}$. For example, for $\alpha =103^{\circ }$ 
\begin{align*}
r_{f}& =0.25,\quad \delta _{f}=154^{\circ },\quad A_{CP}^{f}\approx -0.22 \\
r_{f}& =0.30,\quad \delta _{f}=138^{\circ },\quad A_{CP}^{f}\approx -0.40
\end{align*}

These examples have been selected keeping in view that final state phases $%
\delta_{f, \bar{f}}$ are not too large. For $A^{f, \bar{f}}_{CP}$, we have
used Eq. $\eqref{ccc45}$ neglecting the second order term. An attractive
option is $A_{CP}^{f} = A_{CP}^{\bar{f}}$ for each value of $\alpha$;
although $A_{CP}^{f} \neq A_{CP}^{\bar{f}}$ is also a possibility. $%
A^{f}_{CP} = A_{CP}^{\bar{f}}$ implies $r_{f} = r_{\bar{f}}, \delta_{f} =
\delta_{\bar{f}}$.

Neglecting terms of order $r_{f, \bar{f}}^{2}$, we have 
\begin{align}
A_{CP} \approx \frac{2 \sin \alpha (r_{\bar{f}} \sin \delta_{\bar{f}} -
t^{2} r_{f} \sin \delta_{f} )} {1 + t^{2}} = - \frac{A_{CP}^{\bar{f}} -
t^{2} A_{CP}^{f}}{1 + t^{2}}  \label{cccc1} \\
C \approx - \frac{2 t^{2}}{(1 + t)^{2}} (A_{CP}^{\bar{f}} + A_{CP}^{f})
\label{cccc2} \\
\Delta C \approx \frac{1 - t^{2}}{1 + t^{2}} - \frac{4 t^{2} \cos \alpha}{(1
+ t^{2})^{2}} (r_{\bar{f}} \cos \delta_{\bar{f}} - r_{f} \cos \delta_{f})
\label{cccc3}
\end{align}

Now the second term in Eq. $\eqref{cccc3}$ vanishes and using the value of $%
t $ given in Eq. $\eqref{ccc30}$, we get 
\begin{equation}
\Delta C\approx 0.34\pm 0.06  \label{cccc4}
\end{equation}

Assuming $A_{CP}^{\bar{f}} = A_{CP}^{f}$, we obtain 
\begin{align}
A_{CP} &= - \frac{1 - t^{2}}{1 + t^{2}} A_{CP}^{\bar{f}}  \notag \\
&= (0.34 \pm 0.06) (-A_{CP}^{\bar{f}})  \label{cccc5} \\
C &\approx - \frac{4 t^{2}}{(1 + t^{2})^{2}} A_{CP}^{\bar{f}} \approx -
(0.88 \pm 0.14) A_{CP}^{\bar{f}}  \label{cccc6}
\end{align}

Finally the CP asymmetries in the limit $\delta _{f,\bar{f}}^{T}\rightarrow
0 $ 
\begin{align}
S_{\bar{f}}=S+\Delta S& =\frac{2\text{Im}[e^{2i\phi _{M}}A_{\bar{f}^{\ast }}%
\bar{A}_{\bar{f}}]}{\Gamma (1+A_{CP})}  \notag \\
& =\sqrt{1-C_{\bar{f}}^{2}}\sin (2\alpha _{eff}^{\bar{f}}+\delta )  \notag \\
& =-\sqrt{1-C_{\bar{f}}^{2}}\cos \delta  \label{cccc7} \\
S_{f}=S-\Delta S& =\frac{2\text{Im}[e^{2i\phi _{M}}A_{f}^{\ast }\bar{A}_{f}]%
}{\Gamma (1-A_{CP})}  \notag \\
& =\sqrt{1-C_{f}^{2}}\sin (2\alpha _{eff}^{f}-\delta )  \notag \\
& =\sqrt{1-C_{f}^{2}}\cos \delta  \label{cccc8}
\end{align}

The phase $\delta$ is defined as 
\begin{equation}
\bar{A}_{\bar{f}} = \frac{| \bar{A}_{\bar{f} |}}{| \bar{A}_{f} |} \bar{A}%
_{f} e^{i \delta}  \label{cccc9}
\end{equation}

To conclude:

The final state strong phases essentially arise in terms of $S$-matrix,
which converts an \textquotedblleft $in"$ state into an \textquotedblleft $%
out"$ state. The isospin, $C$-invariance of hadronic dynamics and the
unitarity together with two particle scattering amplitudes in terms of Regge
trajectories are used to get information about these phases. In particular
two body unitarity is used to calculate the final state phase $\delta _{C}$
generated by rescattering for the color suppressed decays in terms of the
color favored decays. In the inclusive version of unitarity, the information
obtained for $s$-wave scattering from Regge trajectories is used to derive
the bounds on the final state phases. In particular, the value obtained for
the final state phases $\delta _{+-}=\delta ^{P}$ $\approx 29^{\circ
}-20^{\circ }$ and $\delta _{00}=\delta ^{C}+\delta ^{P}\approx 20^{\circ
},12^{\circ }$ is found to be compatible with the experimental values for
direct $CP$ asymmetries $A_{CP}(B^{0}\rightarrow \pi ^{-}K^{+},\pi
^{0}K^{0}) $. For\ $B^{0}\rightarrow D^{(\ast )-}\pi ^{+}(D^{(\ast )+}\pi
^{-})$, $B_{s}^{0}\rightarrow D_{s}^{(\ast )-}K^{+}(D_{s}^{(\ast )+}K^{-})$
decays described by two independent single amplitudes $A_{f}$, $A_{\bar{f}%
}^{^{\prime }}$ and $A_{f_{s}},$ $A_{\bar{f}_{s}}^{^{\prime }}$ with
different weak phases viz. $0$ and $\gamma $, equality of phases $\delta
_{f}=\delta _{\bar{f}}^{^{\prime }}$ \ implies, the time dependent CP
asymmetries 
\begin{align}
-\left( \frac{S_{+}+S_{-}}{2}\right) & =\frac{2r_{D_{\left( s\right)
}^{\left( \ast \right) }}}{1+r_{D_{\left( s\right) }^{\left( \ast \right)
}}^{2}}\sin (2\beta _{\left( s\right) }+\gamma )  \label{cccc10} \\
\frac{S_{+}-S_{-}}{2}& =0  \label{cccc11}
\end{align}

An added advantage is that these decays are described by tree graphs.
Assuming factorization, the decay amplitude $A_{f}$ can be determined in
term of the form factors $f_{0}^{B-D}(m_{\pi }^{2})$ and $A_{0}^{B-D^{\ast
}}(m_{\pi }^{2})$. The parameter $r_{D^{\left( \ast \right) }}$ can be
expressed in terms of the ratios of the form factors $f_{D}f_{0}^{B-\pi
}(m_{D}^{2})$/$f_{\pi }f_{0}^{B-D}(m_{\pi }^{2})$ and $f_{D^{\ast
}}f_{+}^{B-\pi }(m_{D^{\ast }}^{2})$/$f_{\pi }A_{0}^{B-D^{\ast }}(m_{\pi
}^{2})$. From the experimental branching ratios, we have obtained the form
factors $f_{0}^{B-D}(m_{\pi }^{2})$ and $A_{0}^{B-D^{\ast }}(m_{\pi }^{2})$
which are in excellent agreement with the prediction of HQET. We have also
determined $r_{D^{\ast }}$. For $r_{D^{\ast }}$ we get the value $r_{D^{\ast
}}=0.017\pm 0.003$. Using this value we get the following bound from the
experimental value of $\frac{S_{+}+S_{-}}{2}$ for $B^{0}\rightarrow D^{\ast
-}\pi ^{+}$ decay: 
\begin{equation*}
\sin (2\beta +\gamma )>0.69
\end{equation*}

Using SU(3), for the form factors for $B_{s}^{0} \rightarrow D_{s}^{\ast -}
K^{+} (D_{s}^{\ast +} K^{-})$ decays, we predict 
\begin{align*}
-\left( \frac{S_{+} + S_{-}}{2} \right) &= (0.41 \pm 0.08) \sin (2\beta +
\gamma) \\
&= (0.41 \pm 0.08) \sin \gamma
\end{align*}
in the standard model.

In section-4, the decays $B\rightarrow \rho ^{+}\pi ^{-}\left( \rho ^{-}\pi
^{+}\right) $ for which decay amplitudes $A_{\bar{f}}$ and $A_{f}$ are given
in terms of tree and penguin diagrams are discussed. We have analyzed these
decays assuming factorization for the tree graph. Factorization implies $%
\delta _{f}^{T}=\delta _{\bar{f}}^{T}$. In the limit $\delta _{f,\bar{f}%
}^{T}\rightarrow 0$, we have shown that 
\begin{align*}
r_{f,\bar{f}}\cos \delta _{f,\bar{f}}& =\cos \alpha \\
r_{f,\bar{f}}^{2}& \approx \cos ^{2}\alpha +A_{CP}^{f,\bar{f}2}\sin
^{2}\alpha
\end{align*}

The first equation has been solved graphically, from which the final state
phases $\delta _{f,\bar{f}}$ corresponding to various values of $r_{f,\bar{f}%
}$ can be found for a particular value of $\alpha $. The upper bound $\delta
_{f,\bar{f}}\leq 30^{0}$ obtained in Section-2, using unitarity and strong
interaction dynamics based on Regge pole phenomonalogy can be used to select
the solutions given in Table-2. Neglecting the terms of order $r_{f,\bar{f}%
}^{2}$, we get using factorization 
\begin{equation*}
\Delta C=0.34\pm 0.06
\end{equation*}

Finally, in the limit $\delta _{f,\bar{f}}^{T}\rightarrow 0$, we get 
\begin{equation*}
\frac{S_{\bar{f}}}{S_{f}}=\frac{S+\Delta S}{S-\Delta S}=-\frac{\sqrt{1-C_{%
\bar{f}}^{2}}}{\sqrt{1-C_{f}^{2}}}
\end{equation*}

With the present experimental data, it is hard to draw any definite
conclusion.

\textbf{Acknowledgement }The author acknowledges a research grant provided
by the Higher Education Commission of Pakistan as a Distinguished National
Professor.

\textbf{Figure Caption:}

Plot of equation $r_{f}\cos \delta _{\left( f\right) }=\cos \alpha $ for
different values of $r.$ For $80^{o}\leq \alpha \leq 103^{o}.$ \ \ Where
solid curve, dashed curve, dashed doted curve, dashed bouble doted and
double dashed doted curve are corresponding to $r=0.1,\ r=0.15,\ r=0.2,\
r=0.25$ and $r=0.3$ respectively.


\begin{thebibliography}{99}
\bibitem{1} J. F. Donoghue et.al. Phys. Rev. Lett. 77, 2187 (1996).

\bibitem{2} M. Suzuki and L. Wofenstein, Phys. Rev. D 60, 074019 (1999).

\bibitem{3} A. Falk et.al. Phys.Rev.D 57, 4290 (1998) hep-ph/9712225.

\bibitem{4} I.Caprini, L.Micer and C.Bourrely, Phys.Rev.D 60,074016 (1999)
hep-ph/9904214.

\bibitem{5} Fayyazuddin, JHEP 09, 055 (2002).

\bibitem{6} Fayyazuddin, Phys. Rev. D70, 114018 (2004).{}

\bibitem{7} M.Gronau and J.L. Rosner, hep-ph/0807.3080 v3.

\bibitem{8} L. Wolfenstein, hep-ph/0407344 v1 (2004); N. Spokvich, Nuovo
Cinento, 26, 186 (1962); K. Gottfried and J. D. Jackson, Nuovo einento 34,
735 (1964); see also, Fayyazuddin and Riazuddin, Quantum Mechanics, Page
140, World Scientific (1990).

\bibitem{9} J. D. Bjorken, Topics in B-physics, Nucl. Physics 11
(proc.suppl.) 325 (1989).

\bibitem{10} M. Beneke, G. Buchalla, M. Neubart and C. T. Sachrajda, Phys.
Rev. Lett, 83, 1914 (1999), Nucl. Phys. B591 313 (2000).

\bibitem{11} C. W. Bauer, D. Pirjol and I. W. Stewart, Phys. Rev. Lett. 87,
201806 (2001) hep-ph/0107002.

\bibitem{12} Particle Data Group, C. Amsler, et.al, Phys. Lett B667,1 (2008).

\bibitem{13} For a review, see for example CP-violation editor: C. Jarlskog,
World scientific (1989); H. Quinn, B physics and CP-violation
[hep-ph/0111174]. Fayyazuddin and Riazuddin, A Modern Introduction to
Particle physics, 2nd edition, world scientific.

\bibitem{14} S. Balk, J. G. Korner, G. Thompson, F. Hussain Z. Phys. C 59,
283-293 (1993).

\bibitem{15} N. Isgur and M. B. Wise Phys. Lett B 232, 113 (1989) Phys.
Lett. B 237, 527 (1990).

\bibitem{16} S. Faller et.al. hep-ph/0809.0222 v1.

\bibitem{17} P. Ball, R. Zweicky and W. I. Fine hep-ph/0412079 v1.

\bibitem{18} G. Duplancic et.al. hep-ph/0801.1796 v2.

\bibitem{19} V. Page and D. London, Phys. Rev. D 70, 017501 (2004).

\bibitem{20} M. Gronau and J. Zupan: hep-ph/0407002, 2004 Refernces to
earlier literature can be found in this ref.

\bibitem{21} Y.Grossman and H.R.Quinn, Phys.Rev.D 58, 017504 (1998);
J.Charles, Phys.Rev.D 59, 054007 (1999); M.Gronau et.al. Phys.Lett B 514,
315 (2001)

\bibitem{22} M. Beneke and M. Neuebert, Nucl. Phys. B675, 333 (2003).
\end{thebibliography}
\end{document}